\newcommand{\refe}[1]{(\ref{#1})}

\documentstyle[aps,twocolumn,psfig]{revtex}	

\begin{document}

\twocolumn[\hsize\textwidth\columnwidth\hsize\csname@twocolumnfalse\endcsname

\title{Geometric Phases and Multiple Degeneracies in Harmonic Resonators}

\author{F. Pistolesi$^{a,*}$ and Nicola Manini$^{b,*,\dag}$} 
 
\address{
$^a$Institut Laue Langevin, B.P. 156, F-38042 Grenoble Cedex 9, France\\
$^b$European Synchrotron Radiation Facility,
B.P. 220, F-38043 Grenoble Cedex, France
}

\date{26 April 2000}
\maketitle

\begin{abstract}
In a recent experiment Lauber {\em et al.}\ have deformed cyclically a
microwave resonator and have measured the adiabatic normal-mode
wavefunctions for each shape along the path of deformation.
%
%
The nontrivial observed cyclic phases around a 3-fold  degeneracy 
were accounted for by Manolopoulos
and Child within an approximate theory.
However, open-path geometrical phases disagree with experiment.
By solving exactly the problem, we find unsuspected extra degeneracies
around the multiple one 
that account for the measured phase changes throughout the path.
It turns out that proliferation of additional degeneracies around 
a multiple one is a common feature of quantum mechanics.
\end{abstract}

] \narrowtext


Geometric phases may show up whenever the system under study depends on
several parameters and is transported adiabatically around a closed path in
parameters space\cite{Berry}.
Such phases have been predicted and observed in many different systems
\cite{Berry,Resta94,Delacretaz,Tycko87,Weinfurter90,Lauber94}.
In particular, the cyclic phases observed by Lauber {\it et 
al.}\cite{Lauber94} for a deformed microwave resonator have been recently
accounted for within an approximate theory \cite{manolopoulos}.
However, the closed-path phases are not the only interesting observable
quantities: for an open-path evolution, the phase of the projection of a
parallel-transported \cite{Anandan87} eigenstate on the initial eigenstate
is equally well
defined \cite{Pancharatnam,Samuel88,Wagh98} whenever these two states are
not orthogonal.
The experiment of Ref.~\cite{Lauber94} measures also such phase
relations for a number of intermediate points along the path.
These phases can be
compared with the theoretical outcome of the calculation of
Ref.~\cite{manolopoulos}.
We observe that the two sets of values disagree at half loop.
The geometrical nature of these phases implies that the approximation of
Ref.~\cite{manolopoulos} is missing the topological structure of the simple
system at hand.

To find what is missing, we push the perturbative expansion of
Ref.~\cite{manolopoulos} to higher order and demonstrate the presence of
unsuspected extra degeneracies which account for the phase discrepancies.
We show that similar ``satellite'' degeneracies close to a multiple
degeneracy appear as a systematic, predictable feature of the Laplace
operator in distorted domains: they can therefore arise in a broad range of
undulatory phenomena, including acoustics, optics and quantum mechanics.

In Ref.~\cite{Lauber94} Lauber {\em et al.}\ follow the adiabatic evolution
of three eigenstates of a microwave resonator driven along a loop in the
two-dimensional space of deformations around the rectangular geometry.
We indicate with ${\bf r} = (r \cos \theta, r \sin \theta)$ the external
parameter representing the displacement of the right upper corner of the
cavity from the rectangular position $(a,b)=(\sqrt{3}, 1)$\cite{sideb:note}
at which the eigenstates $|\psi_1\rangle = \{2,4\}$, $|\psi_2\rangle = \{5,3\}$
and $|\psi_3\rangle = \{7,1\}$ become degenerate [with $\{n_x,n_y\}$ we
indicate the wavefunctions $2\sin(n_x x/a) \sin(n_y y/b)/\sqrt{a b}$ in the
rectangle].
We define the angle $\theta$ so that it goes from $0$ to $2\pi$ following
the path of distortions used in the experiment.
In Fig.~\ref{exp3waves:fig} we report a sequence of adiabatic
eigenfunctions $|\psi_{j}(\theta)\rangle$ from the original pictures of
Ref.~\cite{Lauber94}.


%
\begin{figure}[tbh]
\centerline{
\psfig{file=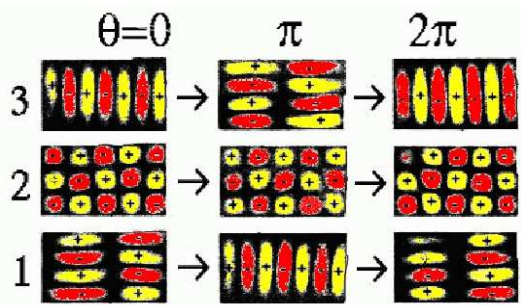,width=7.0cm}
}
\caption{
The observed initial ($\theta=0$), intermediate ($\theta=\pi$) and final
($\theta=2\pi$) eigenstates of the microwave cavity deformed following
adiabatically the path of Ref.~\protect\cite{Lauber94}.
\label{exp3waves:fig}
 }
\end{figure}

The cyclic phase factors [$\gamma_1(2\pi)=-1$, $\gamma_2(2\pi)=+1$
and $\gamma_3(2\pi)=-1$] are easily identified in Fig.~\ref{exp3waves:fig}
from the recurrence of the pattern and the sign changes at $\theta=0$ and
$2\pi$.
Similarly, the sign change of the central state $|\psi_{2}(\theta) \rangle$ at
$\theta=\pi$ tells us that its Berry-Pancharatnam open-path phase factor is
$\gamma_2(\pi)=-1$.

Let us compare now these results with those of the linear
approximation\cite{manolopoulos}. 
The eigenstates of the 2-dimensional Laplacian operator represent the
normal modes of this cavity.
The secular problem in the deformed domain can be conveniently
reduced\cite{manolopoulos} to the secular problem of the following
differential operator 
\begin{equation}
H({\bf r})=
\sum_{i,j} \frac {\partial}{\partial u_i} M_{ij}({\bf u},{\bf r})  
\frac {\partial}{\partial u_j}
+D({\bf u},{\bf r})
\end{equation}
with vanishing boundary conditions in an undistorted rectangular domain for
the transformed variables ${\bf u}$.
This form, given explicitly in Ref.~\cite{manolopoulos}, is particularly
suitable to set up a perturbation theory in $r$
\begin{equation}
	H({\bf r}) 
	= 
	r\, H^{(1)}(\theta) +r^2\, H^{(2)}(\theta)+\dots 
	\quad,
	\label{expansion:eq}
\end{equation}
since all the parametric dependence is now in the functions $M_{ij}({\bf
u},{\bf r})$ and $D({\bf u},{\bf r})$.
At first order it suffices to consider the subspace span by the three
eigenstates of Lauber {\it et al.}'s experiment, and compute the
$3\times 3$ matrices
$
{\cal H}^{(1)}_{ij}(\theta)=\langle \psi_i | H^{(1)}(\theta) | \psi_j\rangle
$.
The form of this matrix, given in Ref.~\cite{manolopoulos}, is
\begin{equation}
	{\cal H}^{(1)}(\theta) 
	= \cos \theta \  F + \sin \theta \  F' \, ,
	\label{Hfirstorder}
\end{equation}
where $F$ and $F'$ are real symmetric numerical matrices.
%
%
In a sufficiently small neighborhood of the 3-fold degeneracy,  ${\cal
H}^{(1)}(\theta)$ accounts for the leading contribution to the
energy shifts.

%

%
\begin{figure}[tbh]
\centerline{ 
\psfig{file=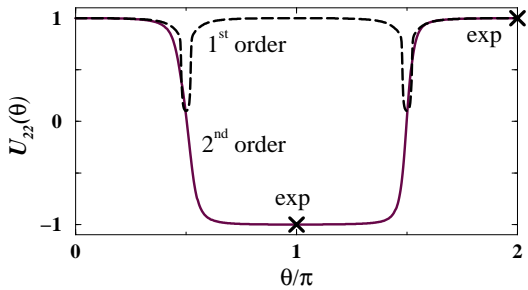,width=7.0cm}
}
\caption{
The projection $U_{22}(\theta)\equiv\langle
\psi_2(0)|\psi_2(\theta)\rangle$ calculated
along the path of Lauber {\it et al.}'s experiment.
Observation\protect\cite{Lauber94}: crosses
(cf.\ Fig.~\protect\ref{exp3waves:fig}).
\label{thwavef:fig}
}
\end{figure}

Manolopoulos and Child \cite{manolopoulos} calculate the (real)
transformation matrix $U^{(1)}_{ij}(\theta)= \langle \psi_i(0)|
\psi_{j}(\theta)\rangle$, which diagonalizes ${\cal
H}^{(1)}(\theta)$ parallel-transporting the eigenstates.
In particular, $U^{(1)}_{ii}(2\pi)$ is the cyclic geometric phase of
state $i$.
Each calculated phase agrees with that observed experimentally.
More generally, for any $\theta$,
$U^{(1)}_{ii}(\theta)/|U^{(1)}_{ii}(\theta)|$ is the observable (open-path)
Berry-Pancharatnam phase factor $\gamma_i$.
These phase factors are therefore represented by the signs of the curves of
Fig.~2b of Ref.\cite{manolopoulos}.
In Fig.~\ref{thwavef:fig} we report $U^{(1)}_{22}(\theta)$ for the
intermediate state $|\psi_2\rangle $.
Near $\theta=\pi$, the disaccord with experiment is particularly evident:
linear theory has $U^{(1)}_{22}(\theta) \simeq +1$, while experimentally
$\gamma_{2}(\pi)=-1$.

\vbox{
\begin{table}[h]
\begin{center}
\begin{tabular}{l|ccc}
	& First order \cite{manolopoulos}
		& Experiment \cite{Lauber94} 	& 2nd order \& BW\\
\hline
$\gamma_2$		&	$+1$	 &	$-1$	& $-1$	\\
$\gamma_{13}$		&	$-1$	 &	$+1$	& $+1$	\\
\end{tabular}
\end{center}
\caption{
Phase factors $\gamma_2$ and $\gamma_{13}$ for the open path $\Gamma_1$
going from $\theta=0$ to $\theta=\pi$ in the space of deformations.
\label{gammas:tab}
}
\end{table}
}

Around $\theta=\pi$, $|U_{11}|$ and $|U_{33}|$ are very small
($\ll 1$): the interesting geometric phase information is contained in the
off-diagonal part of the matrix $U(\pi)$ \cite{ManPis}.
In particular, in this case $ \gamma_{13}(\theta) \equiv U_{13}(\theta)
U_{31}(\theta)/ |U_{13}(\theta) U_{31}(\theta)| $ is the only observable
phase associated to state 1 and 3, since $|U_{13}(\theta)
U_{31}(\theta)|\approx 1$.
 From Fig.~\ref{exp3waves:fig} one can read $U_{13}(\theta)
\approx +1$, and $U_{31}(\theta) \approx +1$. Thus the observed
$\gamma_{13}=+1$.  We have verified that the linear theory of
Ref.~\cite{manolopoulos} gives instead $\gamma_{13}=-1$.

A similar reasoning can be applied to the path $\Gamma_2$ leading back from
$\theta=\pi$ to $\theta=2 \pi$.
The observed and calculated phase changes for this second path
are the same as those for path $\Gamma_1$ going from $\theta=0$ to
$\theta=\pi$: they are summarized in Table~\ref{gammas:tab}.

The failure of the linear approximation in predicting the correct
geometrical phase factors suggests that it misses the topological
structure of degeneracies in parameters space.
In this Letter, we investigate this issue by considering both the next
order in the expansion \refe{expansion:eq} and an exact numerical solution.

 
%
\begin{figure}[tbh]
\centerline{ 
\psfig{file=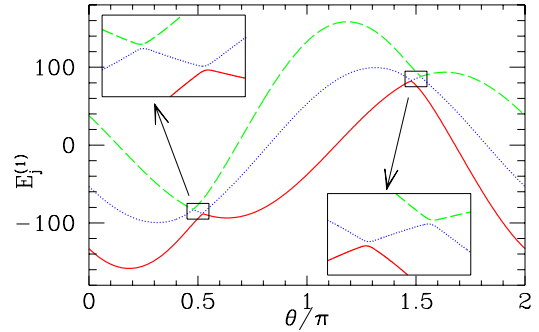,width=7cm,angle=-90}
}
\caption{Linear correction  $E^{(1)}_j(\theta)$ 
[eigenvalues of $H^{(1)}(\theta)$] to the energies of the three states
considered in the text as a function of $\theta$.
}
\label{energy:fig}
\end{figure}

The origin of the problem  can be understood from inspection of
Fig.~\ref{energy:fig}.
Indeed,
close to $\theta =
\pi/2$ and $3\pi/2$, the eigenenergies of ${\cal H}^{(1)}$ get pairwise very
near (the minimum splittings being $\sim 1\%$ of those in the $\theta=0$
direction).
In this region, the linear approximation breaks down for $r$ much smaller
than in the (say) $\theta=0$ direction.
Second-order terms can close the gaps inducing additional degeneracies in
these directions, remaining dominant with respect to third- and higher-order
ones.
It is the smallness of the first-order splitting {\em coefficient} in this
direction $\theta =\pi/2$ and $3\pi/2$ that makes this possible.

We compute therefore the second-order Hamiltonian correction\cite{Landau3}:
\begin{equation}
{\cal H}^{(2)}_{ij} =
\langle \psi_i | H^{(2)} |\psi_j\rangle +
\sum_{k\neq 1,2,3} \frac{
\langle \psi_i | H^{(1)} |\psi_k\rangle
\langle \psi_k | H^{(1)} |\psi_j\rangle}
{E_i-E_k}
\label{h2:eq}
\end{equation}
To calculate explicitly ${\cal H}^{(2)}$, we cutoff the sum of
Eq.~\refe{h2:eq} for $|E_i-E_k| < E_{\rm max}$.
Convergence is achieved including $\sim 10^2$ states.
We have thus obtained the real symmetric numerical matrices $G$, $G'$, and
$G''$ in 
$
{\cal H}^{(2)}(\theta) 
	= 
	G + G'\;\cos 2 \theta + G''\;\sin 2 \theta
$.
%
This second-order contribution adds to the linear Hamiltonian
\refe{Hfirstorder} to constitute the secular problem for the matrix
\begin{equation}
r\,{\cal H}^{(1)}(\theta)+r^2\,{\cal H}^{(2)}(\theta) \, .
\label{Hsum:eq}
\end{equation}
We thus obtain the eigenvalues and eigenvectors of \refe{Hsum:eq} as
analytical functions of $r$ and $\theta$.
Like in the linear case, the transformation matrix $U^{(2)}_{ij}(r,\theta)$
contains all the phase information of the eigenvectors.
Now, at second order, all the geometrical phases, both for the closed loop
and for the open-path, agree with experiment \cite{Lauber94} (cf.\
Tab.~\ref{gammas:tab} and Fig.~\ref{thwavef:fig}).

\begin{figure}[tbh]
\centerline{ 
\psfig{file=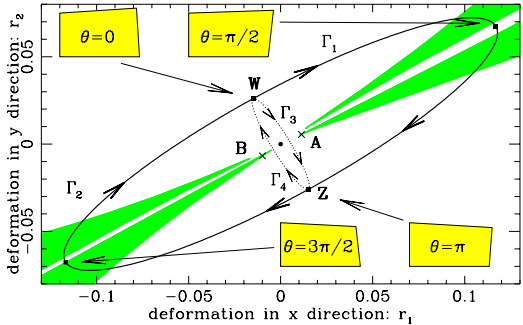,width=7.0cm,angle=-90}
}
\caption{
The path $\Gamma_1+\Gamma_2$ in ${\bf r}$ space 
used in experiment \protect\cite{Lauber94}.
The 3-fold degeneracy sits at the origin.
The four deformed rectangles represent the shape of the
resonator at four representative points along the path.
The two crosses mark the positions of the satellite 2-fold
degeneracies {\bf A} and {\bf B}.  
The non-shaded region is where the linear approximation for the gaps
is correct within 50\%.
\label{exppath:fig}
}
\end{figure}

To elucidate the differences between first and second order,
we show in Fig.~\ref{exppath:fig} the region of
validity of the linear approximation.
The endpoints {\bf W} and {\bf Z} of the open 
path $\Gamma_1$ considered above lie both
well within this region.
In addition, there exists a path $\Gamma_3$ connecting {\bf W} to {\bf Z},
lying all inside
the linear region on the same side of the $3$-fold degeneracy.
For path $\Gamma_3$ the first-order theory is sufficient to predict the
correct phases.
Consider now the loop $\Gamma_1 - \Gamma_3$
($\Gamma_3$ is followed backwards
from {\bf Z} to {\bf W}).
If no degeneracies were present inside, the cyclic phase factors
$\gamma_2^{\Gamma_1 - \Gamma_3}$
would be $+1$ for all states.
Then the open-path phases $\gamma_2^{\Gamma_1}$ and $\gamma_2^{\Gamma_3}$
would have to coincide.
The same applies to the off-diagonal $\gamma_{13}$ \cite{ManPis}.
Different phases in first and second order calculations indicate
additional conical intersections of the second-order Hamiltonian
\refe{Hsum:eq} inside the loop.


%
%


Indeed, we find that, at the points {\bf A} ($r_{\rm \bf A}\simeq 0.0126$,
$\theta_{\rm \bf A}\simeq 1.64$) and {\bf B} ($r_{\rm \bf B}\simeq 0.0120$,
$\theta_{\rm \bf B}\simeq 4.64$) of Fig.~\ref{exppath:fig}, the two lowest
eigenvalues are degenerate.
Knowledge of the phases in the linear region around the threefold
degeneracy at the origin \cite{manolopoulos} together with the presence of
the twofold conical intersections at {\bf A} and {\bf B} accounts for all
(open-path and cyclic) phases.
In particular, on the complete loop $\Gamma_1 +\Gamma_2$, the two sign
changes that $|\psi_1\rangle$ and $|\psi_2\rangle$ acquire encircling {\bf
A} and {\bf B} cancel out: as a fortuitous coincidence, the observed cyclic
Berry phases end up being the same as these obtained in the linear
theory\cite{manolopoulos}.

As well known\cite{VonNeumann} for a generic real Hamiltonian it is
necessary to vary two parameters in order to make a degeneracy occur
accidentally, i.e.\ not on account of symmetry.
In other terms, the codimension of the degeneracies is two.
Thus, in our two-parameter system, 2-fold degeneracies occur at isolated
points.
Therefore, a small readjustment of the two parameters can always compensate
a sufficiently small perturbation to the Hamiltonian: the conical
intersections will only displace slightly upon inclusion of third- and
higher-order corrections of expansion \refe{expansion:eq}.
The fact that the expansion parameter $r$
at the degeneracies is so small indicates that, most likely,
higher-order corrections are indeed smaller
that the $r^2$ term at the degenerate points and even beyond them.
In this case, higher-order terms do not change the qualitative
configuration of degenerate points and do not affect therefore the
geometrical phases calculated at second order.
However, without an upper bound on the higher-order terms, nothing
guarantees that
these terms do not displace significantly or even remove the degenerate
points found at second order.

For this reason, we need to verify with a nonperturbative method the actual
presence and position of the degeneracies.
We thus resort to the numerical technique introduced by Berry and
Wilkinson\cite{BerryW} (BW) to solve the Laplace problem in two-dimensional
domains.
The advantage of this method is to reduce the two-dimensional
differential equation to a one-dimensional integral equation for the normal
derivative of the wavefunction at the boundary, which can be solved on a
discrete mesh.
%
%
We assess the presence of each degeneracy by encircling it with
a loop in the parameters space and calculating the cyclic Berry phases of
the involved states with the BW numerical method.
Specifically, for the problem at hand we have calculated with the BW
technique\cite{NumberN:note} the Berry phases for the (discretized) loop
$\Gamma_1 - \Gamma_3$ in Fig.~\ref{exppath:fig},
obtaining $\gamma_1=-1$, $\gamma_2= -1$, and $\gamma_3=+1$.
Similar procedure also confirms the presence of the degenerate point {\bf B}.

At this stage, one may wonder if satellite degeneracies 
(i.e.\ degeneracies within the range of validity of perturbation
theory, involving irrelevant components on states outside the multiplet)
are present
only in this specific distorted ``quantum billiard'', or if they are a
widespread feature.
Consider, for example, the 3-fold degenerate multiplet $\{n_x,n_y\}=\{1,3\}$,
$\{4,2\}$, and $\{5,1\}$ in the same geometry $a/b=\sqrt{3}$: in this case we
find no satellite degeneracies in the neighborhood of the 3-fold one, where
second-order perturbation theory holds.
This is not surprising, since over the whole range $0<\theta<2\pi$, the
first-order splittings vary much more slowly than those of
Fig.~\ref{energy:fig}.
Thus, when the quadratic term becomes of the same order of the linear term,
also higher-order terms are of comparable size.

If we could establish a general criterion to predict the
occurrence of satellite degeneracies for the general case of a $n$-fold
degenerate multiplet in a billiard of arbitrary side ratio
$a/b$\cite{sideb:note},
then we could easily realize practical classical/quantum systems
with clustered conical degeneracies, and thus intricate patterns of
parallel-transported wavefunctions.
Satellite degeneracies are to be expected whenever (i) the
smallest first-order gap $\delta^{(1)}_{\pi/2}$ near $\theta=\frac\pi 2$ is
much smaller than those $\delta^{(1)}_{0}$ at $\theta=0$, and (ii) the
second-order matrix elements of $|{\cal H}^{(2)}(\pi/2)|\gg
\delta^{(1)}_{\pi/2}$, so that for $r\sim \delta^{(1)}_{\pi/2}/|{\cal
H}^{(2)}_{ij}(\pi/2)|\ll 1$ first and second order terms are of the same
magnitude
and at the same time
higher-orders are negligible 


We first look for the realization of condition (i).
${\cal H}^{(1)}(\theta)$ [Eq.~\refe{Hfirstorder}] determines the
first-order splittings $\delta^{(1)}_\theta$.
The diagonal matrix elements of $F$ determine the linear splittings
$\delta^{(1)}_{0}$ since the off-diagonal elements either vanish or
are much smaller.
The diagonal matrix elements of $F'$ are instead all equal.
Consequently, at $\theta=\pi/2$, the mixing and the splittings
$\delta^{(1)}_{\pi/2}$ of the three states are governed by the off-diagonal
elements of $F'$.
A parity selection rule applies to these matrix elements:
$F'_{\{n_x,n_y\}\{n'_x, n'_y\}}$ is zero when
\begin{equation}
(-1)^{n_x+n'_x}=(-1)^{n_y+n'_y}=1 \ .
\label{parity:eq}
\end{equation}
In addition, the the nonzero elements (setting $a=b$ for the sake
of an estimate) satisfy
\begin{equation}
|F'_{\{n_x,n_y\}\{n'_x, n'_y\}}|
\leq
\frac{
2^{9/2}\; \frac {n_x}{n'_x}\, \frac {n'_y}{n_y} \,
}{	\left[1-\left(\frac {n_x}{n'_x}\right)^2\right]
	\left[1-\left(\frac {n'_y}{n_y}\right)^2\right]
} \ .
\end{equation}
For states widely separated in $n_x - n_y$ plane ($n_x/ n'_x \ll 1$ and
$n'_y / n_y\ll 1$) this coupling becomes very small.
Small linear gaps occur at $\pi/2$ whenever a state $\{n_x,n_y\}$ in the
degenerate multiplet is weakly coupled to the others, i.e. when all the
off-diagonal matrix elements $F'_{\{n_x,n_y\}\{n'_x n'_y\}}$ are much smaller
than $\delta^{(1)}_{0}$.
Indeed in Lauber's multiplet the linear coupling between states $\{5,3\}$ and
$\{7,1\}$ vanishes because of parity \refe{parity:eq},
and the coupling determining
$\delta^{(1)}_{\pi/2}$ is $F'_{\{2,4\}\{7,1\}}\approx 0.9$, compared to
$\delta^{(1)}_{0}\approx 50$.
As we saw, for the multiplet $\{1,3\}$, $\{4,2\}$, and $\{5,1\}$ the ratio
$\delta^{(1)}_{\pi/2}/\delta^{(1)}_{0}$ is much closer to one, and indeed,
no satellite degeneracies appear there.

We come now to condition (ii).
The matrix elements of ${\cal H}^{(2)}(\pi/2)$ follow no selection rule.
We have verified that these matrix elements are large: for instance
$|{\cal H}^{(2)}_{\{n,n+1\}\{n+1,n\}}(\pi/2)|\sim n^3$.
These matrix elements reduce rapidly as a function of the distance 
$(n_x-n'_x)^2+(n_y-n'_y)^2$ between the states.

In conclusion, whenever in a degenerate multiplet one state is near some
states (so that second-order coupling is large) for which selection rule
\refe{parity:eq} makes first order coupling vanish,
and at the same time it is far from all remaining states (so that
$\delta^{(1)}$ is small), one expects satellite degeneracies.
This rule permits, for instance, to decide that the 3-fold multiplet
$\{1,5\}$, $\{4,4\}$, $\{6,2\}$
for $a/b=\sqrt{5/3}$ has a structure of satellite
degeneracies very similar to that of Lauber's experiment, as we then
verified by BW calculation.

An interesting case emerges from this analysis: when all off-diagonal
matrix elements of $F'$ vanish because of selection rule
\refe{parity:eq}, the linear perturbation does not remove the degeneracy at
$\theta=\pi/2$.
In this direction, the second-order term ${\cal H}^{(2)}$ is therefore the
leading responsible for the splitting of the degenerate levels, which
therefore ``kiss'' gently, instead of intersecting conically as usual.
Many two-fold degenerate multiplets satisfy this condition (for example
$\{2,4\}$ and $\{4,2\}$ for $a/b=1$, or $\{2, 3\}$ and $\{6, 1\}$ for $a/b=2$).

In this work, we have accounted for all experimental findings regarding
the eigenstates of Lauber {\it et al.}'s deformed resonator.
We have found that in deformed rectangular resonators
additional degeneracies (with lower multiplicity) may appear
very close to multiple degeneracies.
These extra degeneracies are responsible for the observed pattern of
geometric phases.
These results rely only on the properties of the Laplace
operator:
for this reason they have a wide scope of applicability and are not
restricted to simple microwave resonators.

We thank Dirk Dubbers for suggesting us to investigate this system and for
useful discussions.


\begin{thebibliography}{17}
\bibitem{sideb:note} We fix the length-scale $b=1$, and
  correspondingly the energy unit $b^{-2}$ throughout the paper.
\bibitem{NumberN:note} A good convergence was reached with a mesh of
  300 points on the boundary and $\sim$ 1000 points along the discretized path
  in the parameters space.
\end{thebibliography}

\end{document}